\documentclass[%
reprint,
superscriptaddress,
amsmath,amssymb,
aps,
prl,
]{revtex4-1}

\usepackage{graphicx}% Include figure files
\usepackage{dcolumn}% Align table columns on decimal point
\usepackage{bm}% bold math
\usepackage[mathlines]{lineno}% Enable numbering of text and display math
\immediate\write18{bibtex \jobname}
\newcommand{\STEREO}{\textsc{Stereo }}
\newcolumntype{C}[1]{>{\centering\arraybackslash}m{#1}}
\begin{document}

%\preprint{}

%\title{New constraints on neutron sterile states and on hidden braneworlds with the STEREO experiment}
\title{Searching for Hidden Neutrons with a Reactor Neutrino Experiment: Constraints from the STEREO Experiment}



\makeatletter
\def\@fnsymbol#1{\ensuremath{\ifcase#1 \or \dagger \or \dagger\dagger \or \ddagger \or \ddagger\ddagger \or * \or \mathsection \or \mathsection\mathsection \or ** \or \parallel \or \mathparagraph \else\@ctrerr\fi}}
\makeatother

\newcommand{\MPIK}{\affiliation{Max-Planck-Institut f\"ur Kernphysik, Saupfercheckweg 1, 69117 Heidelberg, Germany}}
\newcommand{\LAPP}{\affiliation{Univ.~Savoie Mont Blanc, CNRS, Laboratoire d'Annecy de Physique des Particules - IN2P3, 74000 Annecy, France}}
\newcommand{\LPSC}{\affiliation{Univ.~Grenoble Alpes, CNRS, Grenoble INP, LPSC-IN2P3, 38000 Grenoble, France}}
\newcommand{\CEA}{\affiliation{IRFU, CEA, Universit\'e Paris-Saclay, 91191 Gif-sur-Yvette, France}}
\newcommand{\UTINAM}{\affiliation{Institut UTINAM, UMR 6213 CNRS, Universit\'{e} Bourgogne--Franche-Comt\'{e}, 25000 Besan\c con, France}}
\newcommand{\LARN}{\affiliation{Department of Physics, University of Namur, 61 rue de Bruxelles, B-5000 Namur, Belgium}}
\newcommand{\ILL}{\affiliation{Institut Laue-Langevin, CS 20156, 38042 Grenoble Cedex 9, France}}

\author{H.~Almaz\'an}\altaffiliation[Present address: ]{Donostia International Physics Center, Paseo Manuel Lardizabal, 4, 20018 Donostia-San Sebastian, Spain}\MPIK
\author{L.~Bernard}\altaffiliation[Present address: ]{Ecole Polytechnique, CNRS/IN2P3, Laboratoire  Leprince-Ringuet, 91128 Palaiseau, France}\LPSC
\author{A.~Blanchet}\altaffiliation[Present address: ]{LPNHE, Sorbonne Universit\'e, Universit\'e de Paris, CNRS/IN2P3, 75005 Paris, France}\CEA
\author{A.~Bonhomme}\MPIK
\author{C.~Buck}\MPIK
\author{P.~del~Amo~Sanchez}\LAPP
\author{I.~El~Atmani}\altaffiliation[Present address: ]{Hassan II University, Faculty of Sciences, A\"in Chock, BP 5366 Maarif, Casablanca 20100, Morocco}\CEA
\author{L.~Labit}\LAPP
\author{J.~Lamblin}\email{jacob.lamblin@lpsc.in2p3.fr}\LPSC
\author{A.~Letourneau}\CEA
\author{D.~Lhuillier}\CEA
\author{M.~Licciardi}\LPSC
\author{M.~Lindner}\MPIK
\author{T.~Materna}\CEA
\author{O.~M\'{e}plan}\LPSC
\author{H.~Pessard}\LAPP
\author{G.~Pignol}\LPSC
\author{J.-S.~R\'eal}\LPSC
\author{J.-S.~Ricol}\LPSC
\author{C.~Roca}\MPIK
\author{R.~Rogly}\CEA
\author{T.~Salagnac}\altaffiliation[Present address: ]{Institut de Physique Nucl\'eaire de Lyon, CNRS/IN2P3, Univ. Lyon, Universit\'e Lyon 1, 69622 Villeurbanne, France}\LPSC
\author{M.~Sarrazin}\email{michael.sarrazin@ac-besancon.fr}\UTINAM\LARN
\author{V.~Savu}\CEA
\author{S.~Schoppmann}\altaffiliation[Present address: ]{University of California, Department of Physics, Berkeley, CA 94720-7300, USA and Lawrence Berkeley National Laboratory, Berkeley, CA 94720-8153, USA}\MPIK   
\author{T.~Soldner}\ILL
\author{A.~Stutz}\LPSC
\author{M.~Vialat}\ILL

% The "\note" macro will give a warning: "Ignoring empty anchor..."
% you can safely ignore it.

% e-mail addresses: only for the corresponding author


\date{\today}

\begin{abstract}

Different extensions of the standard model of particle physics, such as braneworld or mirror matter models, predict the existence of a neutron sterile state, possibly as a dark matter candidate. 
This Letter reports a new experimental constraint on the probability $p$ for neutron conversion into a hidden neutron, set by the \STEREO experiment at the high flux reactor of the Institut Laue-Langevin. The limit is $p<3.1\times 10^{-11}$ at $95 \%$ C.L. improving the previous limit by a factor of 13. This result demonstrates that short-baseline neutrino experiments can be used as competitive passing-through-walls neutron experiments to search for hidden neutrons.

\end{abstract}

%\keywords{}

\maketitle
%\linenumbers
%\section{Introduction}

For many decades, the existence of sterile or hidden particles interacting only gravitationally or very weakly with the known particles of the standard model (SM) has been considered through many theoretical works \cite{mirrorRev,mBim1,mBim2,mir,BB1,BB2,CPDM3,brane2,brane4,brane5,r4,r1,pheno}. They could result in dark matter candidates \cite{mirrorRev,mBim1,mBim2,CPDM3,brane2,brane4,brane5,r4,r1,pheno} or could shed light on primordial cosmology \cite{mirrorRev,BB1,BB2,CPDM3,brane5,r4,r1}. Some of them can be sterile copies of particles of the SM in our usual spacetime \cite{mirrorRev,mBim1,mBim2,mir,BB1,BB2,CPDM3}, allowing for instance for mirror neutrons. Others can be particles from the SM -  in particular neutrons \cite{pheno,PLB,M4xZ2,coupl} - hidden in a parallel braneworld located along an extra dimension in a bulk spacetime \cite{brane2,brane4,brane5,r4,r1,pheno,PLB,M4xZ2,coupl,swap,coupl2}. In the following, hidden neutron will be used as a generic term.

Such models predict that visible neutrons can convert into hidden neutrons and several experiments search for neutron disappearance \cite{mirexp0,mirexp1,mirexp2,mirexp3,mirexp4,mirexp5,manip}. Hidden neutrons could also convert into visible neutrons allowing for neutron disappearance-reappearance experiments. In the last five years, dedicated experiments \cite{npmth,npm,murmur-result,mirexp6} have been developed in order to test those scenarios.
In this Letter, we use the \STEREO experiment \cite{STEREO} installed at the Institut Laue-Langevin (ILL) in Grenoble (France) to derive a new constraint on the neutron-hidden neutron swapping probability, demonstrating that short-baseline neutrino experiments \cite{STEREO,PROSPECT,Neutrino-4,DANSS,Solid} are opportunistic but competitive passing-through-walls neutron experiments. New upper bounds on the coupling parameter between the hidden state and visible state are also inferred.

%\section{Passing-through-walls neutron experiments with a reacteur neutrino detector}

The two-level Hamiltonian $\mathbf{H}$ describing the present problem can be written as \cite{mirrorRev,npm}: 
\begin{equation}
\mathbf{H}=\left( 
\begin{array}{cc}
E_{v} & \varepsilon \,\mathbf{\varkappa } \\ 
\varepsilon \,\mathbf{\varkappa }^{\dagger } & E_{h}%
\end{array}%
\right) ,  \label{Coupling}
\end{equation}%
where $E_{v}$ and $E_{h}$ are the energies in vacuum of the visible and hidden states, $\varepsilon $ is the coupling parameter between both states, and $\mathbf{\varkappa }$ is a unitary matrix whose exact expression depends on detailed physics of
the model but does not change the phenomenology \cite{mirrorRev,pheno}. 

When neutrons travel through a medium, the production rate of hidden neutrons is governed by both the Hamiltonian $\mathbf{H}$ and the neutron-nuclei collision rate $\Gamma $ which writes $\Gamma= v\Sigma _{S}$ with $v$, the neutron velocity and $\Sigma _{S}$ the macroscopic cross section for neutron scattering in the medium.
The collisions act as quantum projection in visible and hidden states but the rate of quantum projection is $\Gamma/2$. The factor 1/2 comes from the fact that collisions project only the visible state. This picture is supported by the full treatment of the density matrix evolution with a Lindblad equation \cite{mir,npmth} from which the swapping probability $p$ at each projection can be derived:
\begin{equation}
p=\frac{2\varepsilon ^{2}}{\left( \Delta E +V_{F}\right) ^{2}+4\varepsilon^{2}+\hbar^2 \Gamma ^{2}/4},  \label{prob}
\end{equation}%
provided that $p \ll 1$ \cite{npmth}, and where $\Delta E=E_{v}-E_{h}$ is the degeneracy-lifting energy difference between visible and hidden states. The Fermi potential $V_{F}$ of the visible neutron in the medium is added to describe the neutron-medium interaction \cite{cs}. For a free neutron, Eq.~\ref{prob} matches with the related time-averaged Rabi probability usually measured in earlier experiments \cite{manip}.

At a macroscopic scale, the number of hidden neutrons produced per unit volume and per unit time is obtained by multiplying $p$ by the volumic rate of projections in the source \cite{npmth}: 
\begin{equation}
S_{h}(\mathbf{r})= p\,\frac{\Sigma_S}{2} \,\Phi _{v}(\mathbf{r}),  \label{eqn:Sh}
\end{equation}
where $\Phi _{v}(\mathbf{r})$ is the visible neutron flux.
A huge number of neutron-nuclei scatterings enhances the swapping probability in contrast to a free motion in vacuum.

Then, hidden neutrons can freely escape the reactor. At a position $\mathbf{r}_d$, the hidden neutron flux is \cite{npmth} 
\begin{equation}  \label{eqn:Phih}
\Phi_{h}(\mathbf{r}_d)=\frac{1}{4 \pi}\int_{\rm Reactor}\frac{S_h(\mathbf{r})}{|\mathbf{r}-\mathbf{r}_d|^2} d^3 r.
\end{equation}

Similarly, the reverse effect allows neutron reappearance in a detector located close to the reactor.
By measuring the neutron flux inside a detection volume shielded from ambient neutrons \cite{npmth}, it is possible to infer the swapping probability or to set an upper limit, provided that the neutron flux $\Phi _{v}$ in the reactor is known. 
The sensitivity of such an experiment mainly relies on the volume of material enhancing the conversion of hidden neutrons, on the neutron detection efficiency and particularly on the ability to avoid as much as possible any background sources. Considering reactor neutrino experiments, neutrino detection is based on the inverse beta decay (IBD) reaction, $\bar{\nu}+p\rightarrow n+e^{+}$, where a delayed-coincidence approach is used with a positron followed by a neutron capture in Gd or Li loaded scintillator \cite{STEREO,PROSPECT,Neutrino-4,DANSS,Solid}. Since these experiments are designed to maximize the neutron detection efficiency, it is natural to explore their use as passing-through-walls neutron experiments.

%\section{The ILL reactor as a hidden neutron source}

\begin{figure}[!bt]
\includegraphics[width=1.0\columnwidth]{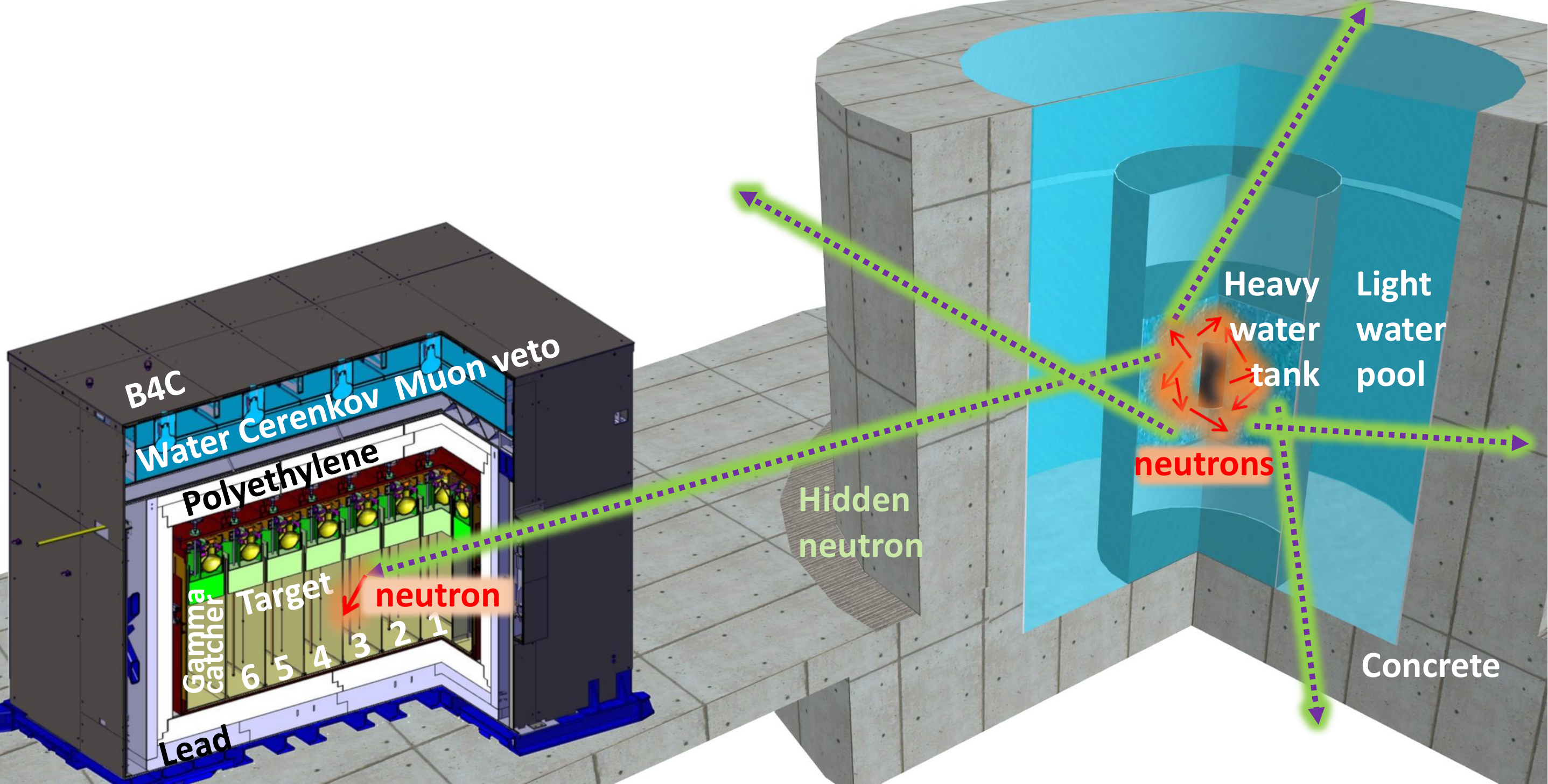}
\caption{The \STEREO detector located at 10~m from the center of the ILL reactor could detect hidden neutrons generated inside the heavy-water tank and regenerated into visible neutrons within the detector (see text for details).}
\label{fig:Detector}
\vspace{-0.5cm}
\end{figure}

The \STEREO experiment (see Fig. \ref{fig:Detector}) was located at 10~m from the center of the ILL High Flux Reactor operated with a 93\% $^{235}$U enriched fuel and heavy water as moderator. The core consists of a single compact fuel element (80~cm high, 40~cm diameter) at the center of a heavy-water tank (1.8~m high, 2.5~m diameter).
The neutron flux map $\Phi _{v}(\mathbf{r})$ within the reactor has been evaluated using convenient numerical computations with Monte Carlo N-Particle transport code MCNP \cite{npm}. 
At nominal power (58.3~MWth), the neutron flux inside the moderator ranges between $10^{14}$ and $1.5 \times 10^{15}$ cm$^{-2}$~s$^{-1}$. It is one of the highest continuous fluxes worldwide thanks to the core design and the low capture cross section of the heavy water which makes the ILL reactor very well suited for hidden neutron conversion. It is largely dominated by thermal neutrons, except in the vicinity of the fuel cylinder, and decreases very fast outside the heavy water tank.
Given these results, we can consider the elastic scattering of thermal neutrons in heavy water, for which the macroscopic cross section is $\Sigma_S^{D_2O}$=0.49 cm$^{-1}$ \cite{cs}, as the dominating hidden neutron conversion mechanism. Neglecting higher energy neutrons and neutron scattering in the light water pool are conservative assumptions.  

Since a simplified geometry was implemented in our simulation, we consider a systematic uncertainty of 20\%, corresponding to the maximum observed discrepancy when comparing with the neutron flux from a full geometry simulation, available only in the median plane \cite{npm}. This systematic uncertainty could be improved by running a precise simulation but would not change our final result which is limited by other uncertainties as shown below. All other systematic uncertainties related to the hidden neutron source, e.g. the time variations due to the fuel evolution, are at the percent level or below \cite{STEREO-norm}.

%\section{Hidden neutron detection}

The target of the \STEREO detector \cite{STEREO} is a $\approx$~2~m$^3$ acrylic aquarium divided in six identical cells filled with Gd loaded liquid scintillator. It is surrounded by an outer crown of 37~cm thickness, namely the gamma catcher, divided in four cells and filled with liquid scintillator without Gd. The gamma catcher ensures a better detection efficiency of the gammas from positron annihilation and neutron capture which can escape the target. For both volumes, the scintillation light is read out from top with a total of 48 photomultipliers.
The gamma catcher vessel is positioned inside a shielding made of borated polyethylene, lead and boron-loaded rubber (B$_4$C) to mitigate gamma and neutron backgrounds. On the top of the shielding, a water Cerenkov detector is installed as muon veto.

In the standard neutrino selection \cite{STEREO-longpaper}, neutron capture events are tagged requiring a reconstructed energy in the whole detector between 4.5 and 10~MeV, the energy of the Gd gamma cascade being $\approx$~8~MeV. The lower bound was chosen to also accept neutrons whose part of gammas escape the detector or depose their energy in nonscintillating components. The request of a delayed coincidence allows one to reject most of the gamma background, important at these energies.
In the case of the hidden neutron search, we expect only single events. We restrict the selection to the 7 to 10~MeV energy window in order to maximize the signal to background ratio.

\begin{figure*}[!ht]
\centering
\vspace{-0.5cm}
\hspace{-0.4cm}
\raisebox{-0.5\height}{\includegraphics[width=0.68\textwidth]{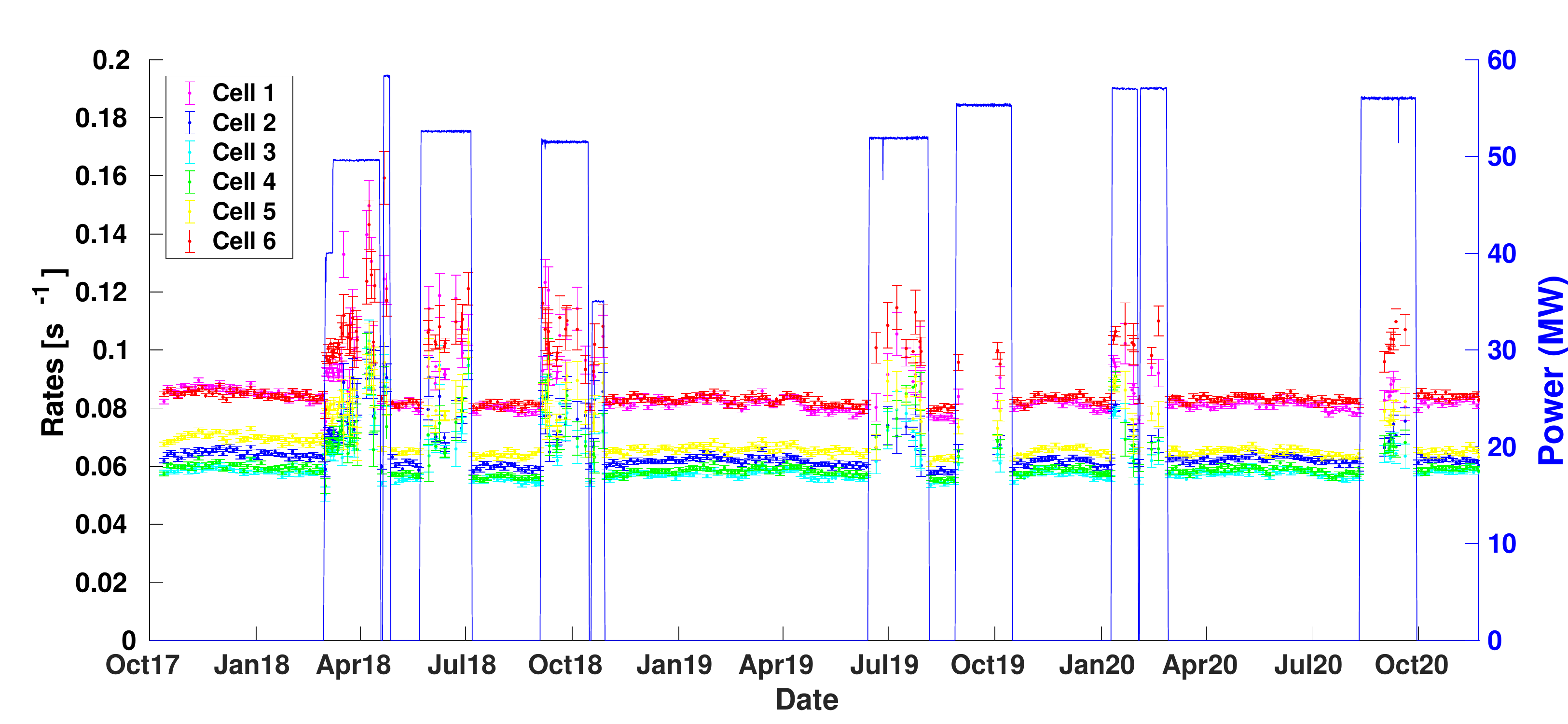}}
\hfill
\hspace{-2cm}
\raisebox{-0.5\height}{\includegraphics[width=0.36\textwidth]{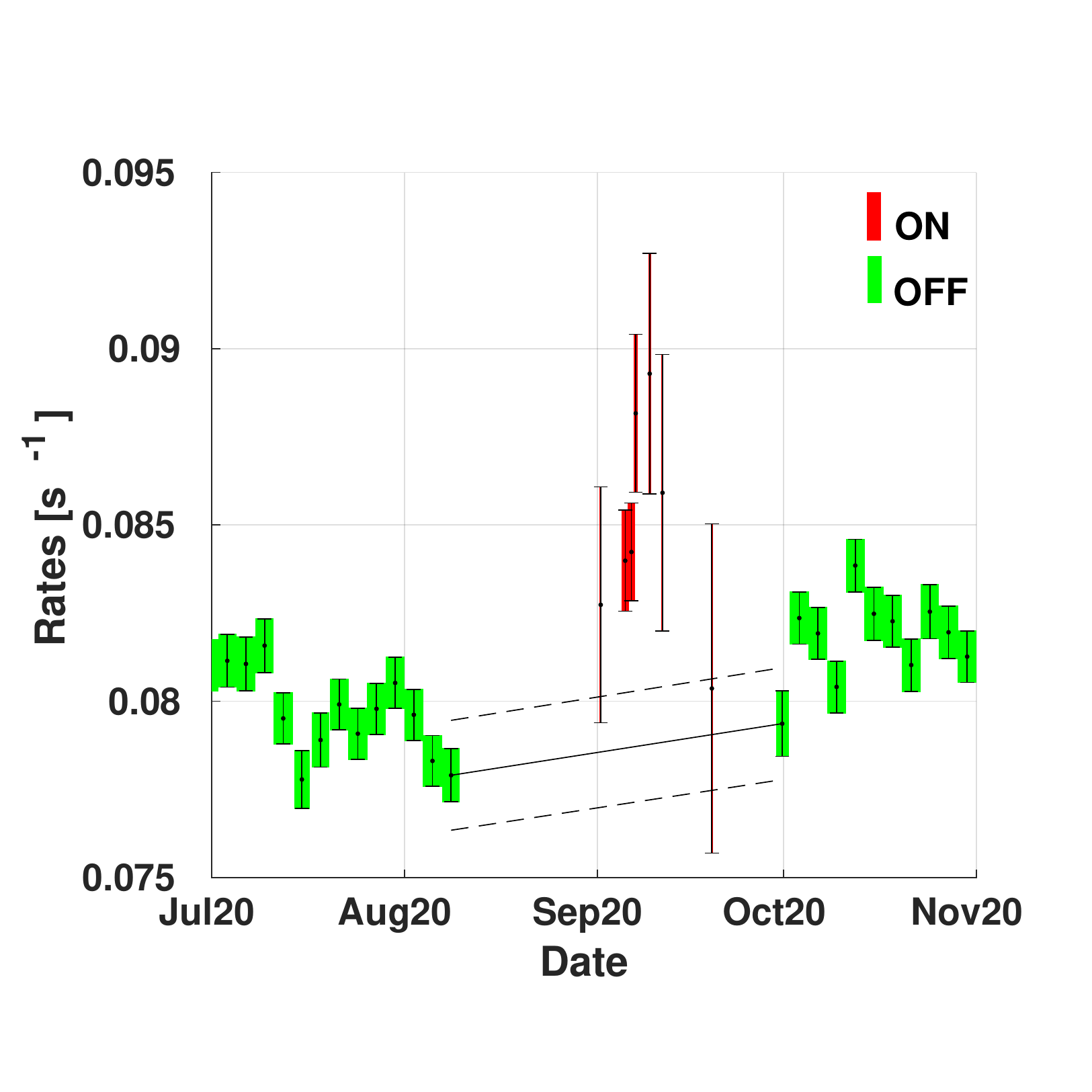}}
\hspace{-0.6cm}
\vspace{-0.5cm}
\caption{(left) Neutron rates $\Gamma$ for each \STEREO cell as a function of time. OFF rates are averaged over three-day bins while one-day bins are used for ON periods for which only quiet periods, per 1~h slot, are kept based on the external BF$_3$ counter monitoring the ambient neutron background. In blue (right axis), the reactor power is also indicated. Right: enlargement of the most quiet period in 2020 for cell 1. The continuous and dashed lines correspond to the linear interpolation for the OFF subtraction and the associated uncertainty. The width of the green and red bands is proportional to the effective time of each bin. }
\label{fig:Rates}
\end{figure*}

Even if \STEREO profits from 15 m.w.e. overburden from the building and a water transfer channel of the reactor, cosmic induced events constitute a significant part of the neutron background. To reject them, we require no other event in the detector nor in the muon veto in a time window $\pm$400~$\mu$s around the neutron events. The time window size has been optimized to maximize the signal to background ratio. This anticoincidence selection generates a dead time of about 50\% during ON periods. The precision of the dead time correction is at the percent level.  

The regeneration of hidden neutrons in the detector can happen either via elastic scatterings in the materials of the detector, mainly the liquid scintillator, or directly via captures on gadolinium. The former process is more probable, $\Sigma_S^{\rm scint}$=1.90 cm$^{-1}$ \footnote{Computed using MCNP.} compared with $\Sigma_C^{\rm Gd}$=0.33 cm$^{-1}$  \footnote{Computed from the Gd fraction and abundances in the scintillator.} but it has a slightly lower detection efficiency since regenerated thermal neutrons can escape the target volume or be captured on hydrogen. 
We denote $\epsilon^n_i(\mathbf{r}_d)$, the probability of a thermal neutron regenerated at the position $\mathbf{r}_d$ to be detected in the cell $i$ and $\epsilon^\gamma_i(\mathbf{r}_d)$, the probability of a neutron capture at the position $\mathbf{r}_d$ to have a vertex reconstructed in the cell $i$.

The detection rate in the target cell $i$ can be written as an integral over the detector:

\begin{equation}\label{eqn:Gammai}
\Gamma_{i}=\int_{\rm det}p \bigg( \frac{\Sigma_S^{\rm scint}}{2}  \epsilon^n_i(\mathbf{r}_d)+ \frac{\Sigma_C^{\rm Gd}}{2}  \epsilon^\gamma_i(\mathbf{r}_d)\bigg) \Phi_h(\mathbf{r}_d) d^3 r_d.
\end{equation}

To simplify the computation, the integral is replaced by a Riemann sum over the detector cells. The cell thickness, 37~cm, is small enough compared with the distance to the core to assume a constant hidden neutron flux within each cell. 

Detection efficiencies have been computed using the \STEREO \textsc{Geant}4 simulation code which has been validated at the percent level using gamma and neutron calibrations \cite{STEREO-longpaper}. Particularly, the use of the FIFRELIN code significantly improved the Gd gamma cascade simulation \cite{STEREO-Fifrelin}. For $\epsilon^n_i(\mathbf{r}_d)$, thermal neutrons have been generated in the whole detector, including the shielding. Indeed, as shown in Ref. \cite{murmur-result}, the shielding materials, and particularly the lead, can enhance the hidden neutron sensitivity. In our case, it appears that the lead is too far from the target ; all regenerated neutrons are captured in the boron-loaded polyethylene. The neutron detection efficiency of a given cell is significant only when the neutron is regenerated in this cell (between 29.9\% and 33.4\% depending on the cell) or in the adjacent cells (between 1.6\% and 4.4\%).
To compute $\epsilon^\gamma_i(\mathbf{r}_d)$, we selected the Gd captures from the same simulation. As expected, the values are slightly higher (between 32.1\% and 35.8\% for the vertex cell, between 3.6\% and 4.5\% for the adjacent cells).  The relative uncertainties on the efficiencies comprise between 1\% and 3\%.

%\section{Measured rates}

The \STEREO experiment has been taking data between 2016 and 2020 \cite{data}.
% However, the whole period is not suited for the hidden neutron search. During phase-I, detector defects (repaired for the next phases) prevented an accurate simulation of the absolute efficiency of all cells. The computed detection efficiencies hold only for phase-II and phase-III. Moreover, The neutron background at the \STEREO location varies a lot, by more than two orders of magnitude, depending on the running conditions of the neighboring experiments.
The neutron background at the \STEREO location depends on the running conditions of the neighboring experiments.
A BF$_3$ counter located on top of the muon veto monitored the neutron rate outside the shielding. During ON periods, the measured rates range between a few neutrons per second to a few hundred neutrons per second and are strongly correlated with the neutron rates in the target cells. Thus, we only consider periods where the BF$_3$ counting rate, averaged per 1 h slot, is below 5 neutrons per second, corresponding to the quietest observed periods.  The use of an external counter to clean the data avoids biasing the analysis. 

Fig. \ref{fig:Rates} shows the resulting neutron rates $\Gamma$ for each cell as a function of time. 
The ON periods are clearly visible with higher and fluctuating rates while the rates are almost constant during OFF periods in between. Cells 1 and 6 present higher rates because they are less shielded than the center cells.
Two main conclusions can be drawn. Firstly, the ON events cannot all correspond to hidden neutrons because their rate should depend only on the reactor power which is almost constant during operation. Secondly, we can take advantage of OFF periods to measure the reactor-operation-independent background and subtract it from ON rates. To that end, we use a linear interpolation between periods of three days before and after each ON cycle. In addition to the statistical uncertainties of the OFF rates, we have to consider a systematic uncertainty to cover a variation of the background during the ON period. This uncertainty can be estimated by testing the subtraction procedure over the whole OFF dataset. The standard deviation of the residuals after subtraction, which includes both statistical and systematic uncertainties, is 2\% of the OFF rates. The three-day binning has been chosen to minimize this uncertainty.

After the OFF subtraction, the rates were the lowest in the beginning of September 2020, when we could benefit from several days with the neighboring experiments not running. For cell 1, the lowest measured ON-OFF rate is $\Gamma_1^{\rm ON-OFF}=(5.3 \pm 2.1)  \times 10^{-3} $~s$^{-1}$. Other cells present similar or higher ON-OFF rates but the hypothetical hidden neutron rate decreases with solid angle. Thus, cell 1, which is the closest cell to the reactor, will provide the best limit. 
Considering that $\Gamma_1^{\rm ON-OFF}$ is still dominated by background events, we derive an upper limit on $p$ from Eqs. (\ref{eqn:Sh})-(\ref{eqn:Gammai}), taking into account the reactor power in the relevant period:
\begin{equation}
p<3.1\times 10^{-11}\ \ (\text{at }95\text{ }\%\text{ C.L.}).  \label{pbound}
\end{equation}%

We can also compute  (see Table~\ref{table:lim_contrib}) what would be the limit by minimizing the value of each measured rate or uncertainty independently (fourth column) or all together except one (last column), keeping the same ON exposure (cell 1 during one day, limited by the typical downtime of neighboring experiments), but allowing longer OFF exposure.

\begin{table}[!h]
\caption{Impact of each measured rate and uncertainty on the limit\label{table:lim_contrib} keeping the same ON exposure (cell 1 during one day) }
\begin{tabular}{|C{2.3cm}C{1.6cm}C{1.4cm}|C{1.2cm}|C{1.4cm}|}
\hline
Quantity $Q_i$ \tt & Current Value & Min Value &  Limit if $Q_i$=Min ($ 10^{-11}$) & Limit if $Q_{j \ne i}$=Min ($ 10^{-11}$) 
\tabularnewline
\hline 
$\Gamma_1^{\rm ON-OFF}  ($s$^{-1})$ & $5.3 \times 10^{-3} $ & 0 & 1.9 &  2.5 \tabularnewline
$\Gamma_1^{\rm OFF} ($s$^{-1})$ &$78.7 \times 10^{-3}$ & 0 & 2.7 &  1.9 \tabularnewline
 $\Delta \Gamma_1^{\rm OFF} ($s$^{-1})$ & $ 1.6 \times 10^{-3}$ & 0 & 2.9 & 0.2 \tabularnewline
 $\Delta \Gamma_1^{\rm ON} ($s$^{-1})$ & $1.4 \times 10^{-3}$ & $2.3 \times 10^{-5}$ & 3 &  0.2 \tabularnewline
 $\Delta \Phi_v / \Phi_v$ & $20\% $ & $1\%$ & 3  & 0.2 \tabularnewline
 Other syst.  & $4.2\%$ & $1\%$ & 3.1 &  0.2 \tabularnewline
 \hline 
\end{tabular}
\end{table}

Table~\ref{table:lim_contrib} shows that the main contribution is due to the $\Gamma_1^{\rm ON-OFF} $ value, i.e. the reactor induced background. The OFF background, $\Gamma_1^{\rm OFF}$, has also a non-negligible impact on the limit. 
ON statistics only contribute for a very small fraction. Using a larger ON dataset would not improve the result ; a one-day binning is sufficient. It is better to select a short period with the lowest reactor induced background. 
As a consequence, this result is limited by the reactor induced background and, in spite of the 15 m.w.e. overburden, by the cosmic-ray induced background. In the case of a background free experiment, the same exposure would give a limit of $2\times 10^{-12}$.
If only the reactogenic background is null, the limit would be $1.9\times 10^{-11}$ .

\begin{figure}[!bt]
\centering
\vspace{-0.5cm}
\includegraphics[width=1.0\columnwidth]{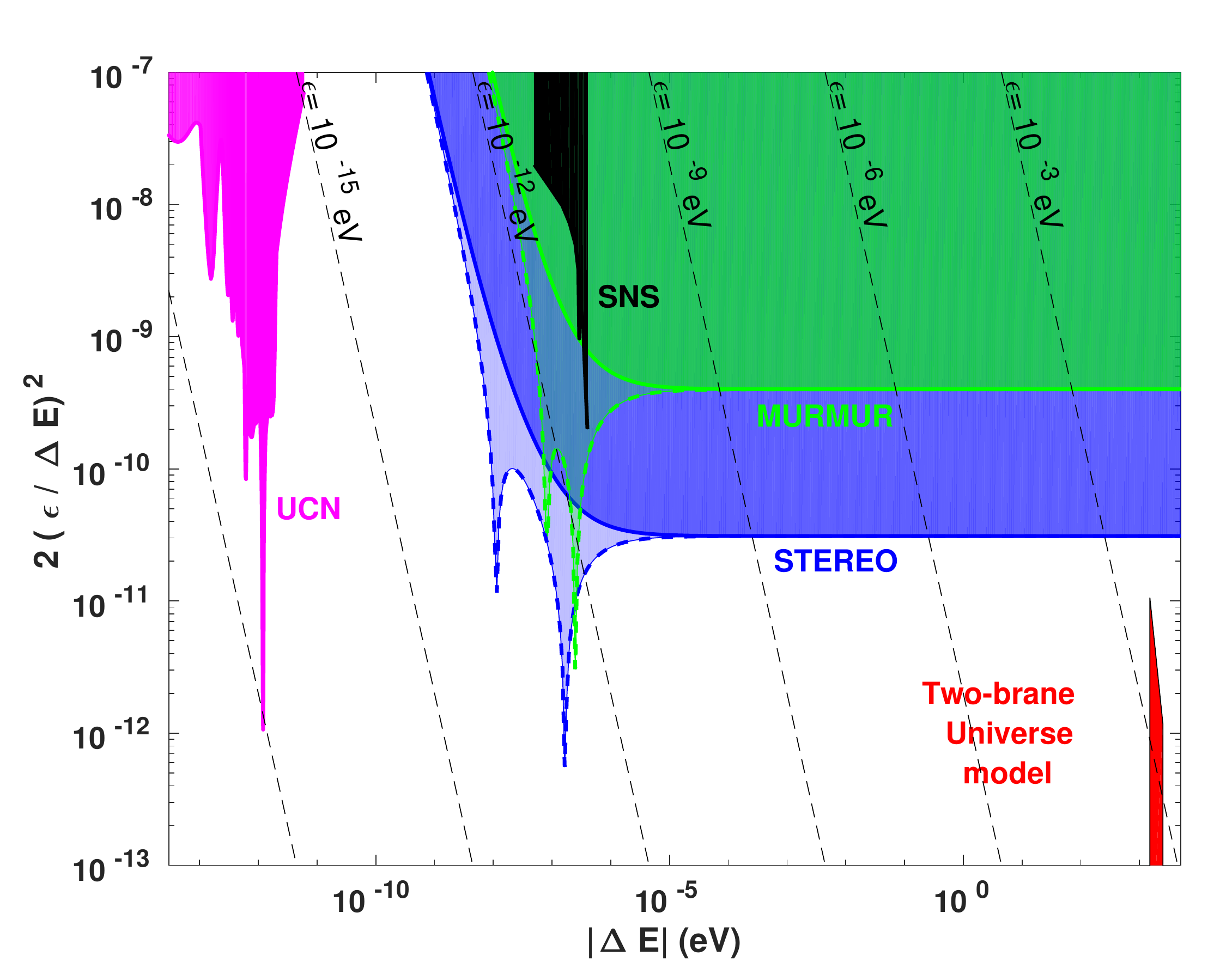}
\vspace{-0.5cm}
\caption{Exclusion contour from \STEREO (blue), MURMUR (green), ultra cold neutrons experiments (magenta) and cold neutron experiment at SNS (black). The dark (light) green and blue contours correspond to the $\Delta E>0$ ($\Delta E<0$) case. The two most sensitive points --- $\Delta E$ = 11 neV and 167 neV --- result from the Fermi potentials of the scintillator and D$_2$O, respectively. The expected region corresponding to the two-brane universe model is also plotted (red). The ratio $2(\epsilon/\Delta E)^2$ corresponds to the Eq.\ref{prob} limit for high values of $\Delta E$.}

\label{fig:bounds}
\end{figure}

The limit [Eq. (\ref{pbound})] is $13$ times better than the previous one obtained with a dedicated experiment, MURMUR \cite{murmur-result}. 
The main reasons for this improvement are the source configuration and a factor hundred lower counting rate per volume unit, thanks to a better shielding. 
 Figure~\ref{fig:bounds} shows the corresponding exclusion contour in the ($\varepsilon $,$\Delta E$) parameter space, obtained using Eq.~(\ref{prob}), compared with the contours of the MURMUR experiment \cite{murmur-result}, ultra cold neutrons (UCN) experiments \cite{mirexp1,mirexp2, mirexp3,mirexp4,mirexp5}, and a cold neutron experiment at the Spallation Neutron Source SNS \cite{mirexp6}. 
%While $\Delta E$ should be zero or close to zero in the context of the original mirror matter approaches \cite{mirrorRev}, it could reach larger values in some scenarios with bimetric gravity \cite{mBim1,mBim2}.
In braneworld approaches, $\Delta E$ naturally merges with the difference of gravitational potential energies felt by the neutron in each brane \cite{pheno,swap}. Using recent data \cite{Planck}, the expected value is around $\Delta E=2$ keV. It can also be shown that the maximum expected value of the coupling parameter is $\varepsilon =2.9$ meV if the brane energy scale equals the Planck energy \cite{coupl,coupl2}.
With a new experimental upper limit $\varepsilon(\Delta E=2$ keV$) =7.9$ meV, the \STEREO experiment gets close to the expected values.

The present work justifies the relevance for short-baseline neutrino experiments to test neutron physics beyond the standard model in the quest for hidden sectors. This approach is very competitive compared with dedicated experiments \cite{murmur-result} or with other kinds of experiments related to mirror matter \cite{mirexp0,mirexp1,mirexp2,mirexp3,mirexp4,mirexp5,mirexp6}. 
The three parameters to be optimized are the hidden neutron flux (i.e., the neutron flux within the reactor, its spatial extension, and the material content since the key parameter is the total number of neutron scatterings per second), the distance between the core and the detector, and last but not least, the level of background. For the first two, the ILL site turns out to be quite optimal. The large ratio of scattering to absorption cross sections in heavy water significantly increases the mean number of collisions per neutron before its capture (2 orders of magnitude difference between heavy water and water). However, the level of background could be further improved either with a better shielding against reactogenic and cosmogenic backgrounds or using a site with less background. A detector with a better neutron discrimination, as, e.g., PROSPECT \cite{PROSPECT} and SoLiD \cite{Solid} detectors with Li-loaded scintillators, could also help to improve the result.

\begin{acknowledgments}
This work received funding from the French National Research Agency (ANR) within the Project No. ANR-13-BS05-0007. The authors are grateful for the technical and administrative support of the ILL for the installation and operation of the \STEREO detector. We further acknowledge the support of the CEA, the CNRS/IN2P3, and the Max Planck Society.\\
\end{acknowledgments}

%\bibliography{FirstResultsSTEREO}

\end{document}